\documentclass[12pt]{article}
\usepackage{amssymb,amsfonts,amsmath}
\usepackage{graphicx} 
\usepackage{indentfirst}
\usepackage{fancybox}

\parskip=\medskipamount

\newcommand{\sect}[1]{\setcounter{equation}{0}\section{#1}}

\newcommand{\be}{\begin{equation}}
\newcommand{\ee}{\end{equation}}
\newcommand{\bea}{\begin{eqnarray}}
\newcommand{\eea}{\end{eqnarray}}

\def\double #1{#1{\hbox{\kern-2pt $#1$}}}

\newcommand{\ket}[1]{\, |#1\rangle}
\newcommand{\bra}[1]{\langle #1 |\,}

\begin{document}

\begin{center}
{\Large \bf  On general properties of Lorentz invariant formulation} \\
\vspace{2mm}
{\Large \bf   of noncommutative quantum field theory}  \\

\end{center}

\begin{center}

{\large  
Sami Saxell\footnote{sami.saxell@helsinki.fi}
} \\
\vspace{5mm}

\footnotesize{
{\it Department of Physics\\
University of Helsinki and Helsinki Institute of Physics\\
P.O. Box 64, FIN-00014, Finland
}}  
\end{center}

\begin{abstract}
\baselineskip=14pt
\noindent
We study general properties of certain Lorentz invariant noncommutative quantum field theories proposed 
in the literature.
We show that
causality in those theories does not hold, in contrast to the canonical
noncommutative field theory with the light-wedge causality condition. This is the consequence of the infinite nonlocality of the theory
getting spread in all spacetime directions.
We also show that the time-ordered perturbation theory arising from the Hamiltonian formulation of noncommutative quantum field theories remains inequivalent to the covariant perturbation theory with usual Feynman rules even after restoration of Lorentz symmetry.

\end{abstract}
\vspace{1cm}

\vfill

\newpage

\sect{Introduction}

There are strong arguments that close to the Planck scale the spacetime manifold should be replaced by noncommutative (NC) structure, which arises through quantum and gravitational effects \cite{Doplicher:1994zv,Doplicher:1994tu}. Field theory on such spacetimes has been an active field of study during recent years. 
The basic object in NC field theory is the nonvanishing commutator of the spacetime coordinates
\begin{eqnarray}
\left[\hat x^\mu , \hat{x}^\nu\right]=i\hat\theta^{\mu\nu}.
\label{commutator}
\end{eqnarray}
The right hand side of this equation can be interpreted in different ways. By considering the effective field theory arising from open string dynamics on a brane in the presence of a constant antisymmetric background field, one obtains a NC spacetime of this type \cite{Seiberg:1999vs}. In this case the r.h.s. is a constant parameter that is related to the background field. Such a constant parameter provides directionality into the spacetime and while it maintains translational invariance, the Lorentz symmetry is broken into the stability group of $\theta^{\mu\nu}$ \cite{AlvarezGaume:2001ka, AlvarezGaume:2003mb}.
This type of NC we will refer to as the canonical noncommutativity.
Due to the connection with string theory, this type of noncommutativity has been studied very extensively. However the violation of Lorentz invariance is a serious drawback leading to effects such as vacuum birefringence \cite{Jaeckel:2005wt}, that are difficult to reconcile with experimental results.

Another option is to consider the r.h.s. as a tensorial operator which commutes with the coordinates. By considering measurements combining principles of classical general relativity and quantum mechanics, Doplicher, Fredenhagen and Roberts (DFR) were led to a spacetime of this type \cite{Doplicher:1994zv,Doplicher:1994tu}. DFR showed that a state which minimizes the uncertainty in time and space in a chosen Lorentz frame corresponds to integrating the tensor $\theta^{\mu\nu}$ over spatial rotations. Such a prescription leads to rotational but not Lorentz invariant theory. From the operator valued NC parameter one obtains the canonical case by choosing an eigenstate of the operator, so that the r.h.s. of Eq. (\ref{commutator}) is replaced by the eigenvalue.

In order to obtain complete Lorentz invariance, Carlson, Carone and Zobin \cite{Carlson:2002wj} constructed NC field theory with integration over all values of $\theta$. To make such an integral convergent they introduced a Lorentz invariant weight function $W(\theta)$. The exact form of the weight function is unknown and in this kind of theory noncommutative physics is parametrized by Lorentz invariant quantities such as
\begin{eqnarray}
\langle\theta^2\rangle=\int d^6\theta W(\theta)\theta^{\mu\nu}\theta_{\mu\nu}.
\label{}
\end{eqnarray} 
These theories allow for the scale of noncommutativity to be low enough to be detectable in future experiments, without leading to disastrous Lorentz violating effects \cite{Carlson:2002wj, Haghighat:2004gd, Carone:2006jz, Ettefaghi:2007cc}.
Lorentz invariance also allows one to consistently take the particles in representations of the full Poincar\'e group without need to refer to the twisted Poincar\'e symmetry that the canonical NC theories possess \cite{Chaichian:2004za}. For another approach to Lorentz symmetry in NC field theories, see \cite{Banerjee:2004ev}.

The purpose of this paper is to study basic properties of Lorentz invariant NC field theory, such as unitarity and causality to see whether such a theory is really valid as a quantum field theory.
Our main result concerns (micro)causality in Lorentz invariant NC field theory. 
In the canonical case the light-cone causality condition is known to be modified. For example for a space-like noncommutativity parameter, $\theta^{\mu\nu}\theta_{\mu\nu}<0 $ one can choose a frame where $\theta^{0i}=0$ and there is noncommutativity only between $x^2$ and $x^3$ -directions. Then events inside the light wedge $(x^0)^2-(x^1)^2>0$ are causally connected implying instantaneous propagation in the $(x^2,x^3)$-plane . The light-wedge causality condition corresponds to the reduced symmetry group, $O(1,1)\times SO(2)$, of such spacetime \cite{AlvarezGaume:2001ka, AlvarezGaume:2003mb}.

In the Lorentz invariant case the symmetry group is again the whole Lorentz group allowing for the possibility of a light-cone causality condition. However, since the symmetrization procedure includes integration over all possible values of the NC parameter, nonlocality spreads in all directions 
and we may well expect that causality will be completely lost in this case. This indeed turns out to be the case. 
In the Lorentz invariant NC theory the situation is even more severe than in the case of canonical noncommutativity where the light-wedge causality condition can still be maintained.
Microcausality is of utmost importance due to the physically accessible implications of it such as dispersion relations \cite{Liao:2002fn,Chaichian:2003hx}. The existence of dispersion relations is the cornerstone to derive an important exact result, the analogue of the Froissart-Martin bound on the high-energy behaviour of the total cross section in NC QFT \cite{Tureanu:2006ct}. 
The acausal effects that arise from time-space noncommutativity make the theory inconsistent as was shown in \cite{Seiberg:2000gc} by studying the scattering of wave packets in canonical NC field theory with $\theta^{0i}\not=0$.

Our other result is on inequivalence of time-ordered perturbation theory (TOPT) and Lorentz
covariant perturbation theory, which is closely related to the question of unitarity in NC field theory.
In commutative spacetime both perturbation theories are equivalent, while in the case of canonical NC spacetime the equivalence holds only if noncommutativity is restricted to the space directions i.e. $\theta^{0i}=0$ \cite{Bahns:2002vm, Liao:2002xc}. In the case of time-space noncommutativity, unitarity is known to be violated in canonical NC quantum field theory if one uses covariant perturbation theory \cite{Gomis:2000zz}. However, in this case the usual covariant perturbation theory written in terms of Feynman propagators cannot be derived from TOPT in which unitarity is manifest \cite{Bahns:2002vm, Liao:2002xc}. In Lorentz invariant NC theory, unitarity was shown to be valid at least in the lowest order in a simple model considered in \cite{Morita:2003vt}, where the calculation was based on covariant perturbation theory. This result leads to the question whether the time-ordered and covariant perturbation theories could be equivalent after restoring the Lorentz symmetry 
despite the fact that the symmetrization procedure itself includes integration over all values of theta, including those with time-space noncommutativity. In this paper we show that this question has a negative answer.

The paper is organized as follows. In section two we review the Lorentz invariant formulation of NC field theory based on the DFR algebra.
In section 3 we study the commutator of two observables from which we conclude the violation of light-cone causality in Lorentz invariant NC theory. In section 4 we calculate a simple tree level Feynman diagram in TOPT and covariant perturbation theory in order to demonstrate their inequivalence. Section 5 is for discussion and conclusions.


\sect{The DFR algebra and Lorentz invariant NC QFT}

By considering uncertainty relations arising from quantum mechanics and general relativity in a position measurement the authors of \cite{Doplicher:1994zv} were led to propose 
a noncommutative algebra for the spacetime coordinates, given by
\footnote{In \cite{Carlson:2002wj} the authors started with the Lorentz invariant NC spacetime of Snyder \cite{snyder} and obtained Poincar\'e invariant spacetime by taking a certain continuum limit. The connection between their spacetime and the DFR formulation was elaborated in \cite{Kase:2002pi}.}:
\begin{eqnarray}
\left[\hat x^\mu , \hat{x}^\nu\right]=i\hat\theta^{\mu\nu},\nonumber\\ 
\left[\hat x^\mu , \hat \theta^{\nu\rho} \right]=0, \\ 
\left[\hat \theta^{\mu\nu} , \hat\theta^{\rho\sigma} \right]=0\nonumber,
\label{algebra}
\end{eqnarray}
which we refer to as the DFR algebra.
The last equation is obtained from the first two by requiring the Jacobi identity to hold.
This algebra is closed under Lorentz transformations.
Next we define the operator field in the DFR algebra in the Weyl representation
\begin{eqnarray}
\hat\phi(\hat x, \hat \theta)= \frac{1}{(2\pi)^4}\int d^4p \; d^6\sigma \;\tilde \phi(p,\sigma) e^{ip_\mu\hat x^\mu+i\sigma_{\mu\nu}\hat \theta^{\mu\nu}}.
\label{ncfield}
\end{eqnarray}
Here $p_{\mu}$ and $\sigma_{\mu\nu}$ are real variables.
The difference from the canonical NC case is that now in general the fields depend also on the operator $\hat\theta$ since it is an element of the spacetime algebra.

Because of the last equation in (2.1), the components of $\hat\theta^{\mu\nu}$ can be diagonalized simultaneously.
Then for any eigenstate of $\hat \theta$:
\begin{eqnarray}
\bra{\theta}\hat\phi(\hat x, \hat \theta) \ket{\theta}=
\bra{\theta} \frac{1}{(2\pi)^4}\int d^4p  \; d^6\sigma \; \tilde \phi(p,\sigma) e^{ip_\mu\hat x^\mu+i\sigma_{\mu\nu}\hat \theta^{\mu\nu}} \ket{\theta}\nonumber\\
= \frac{1}{(2\pi)^4}\int d^4p \; d^6\sigma \; \tilde \phi(p,\sigma) e^{ip_\mu\hat x^\mu+i\sigma_{\mu\nu} \theta^{\mu\nu}}\langle\theta\ket{\theta}
=\hat\phi(\hat x,\theta)\langle\theta\ket{\theta}.
\label{}
\end{eqnarray}

The Weyl symbol corresponding to the field operator is defined by
\begin{eqnarray}
\phi( x, \theta)= \frac{1}{(2\pi)^4}\int d^4p \; d^6\sigma \; \tilde \phi(p,\sigma) e^{ip_\mu x^\mu+i\sigma_{\mu\nu} \theta^{\mu\nu}}.  
\label{}
\end{eqnarray}
The Weyl symbol provides a map from the operator algebra to the algebra of functions equipped with a star-product,
via the Weyl-Moyal correspondence
\begin{eqnarray}
\hat \phi  (\hat x,\hat \theta) \hat \psi (\hat x,\hat \theta) \leftrightarrow \phi(x,\theta)*\psi(x,\theta).
\label{}
\end{eqnarray}
The star-product turns out to be the same as in the canonical case:
\begin{eqnarray}
\phi(x,\theta)*\psi(x,\theta)=\phi(x,\theta)e^{\frac{i}{2}\overleftarrow\partial_\mu\theta^{\mu\nu}\overrightarrow\partial_\nu}\psi(x,\theta).
\label{star}
\end{eqnarray}
Due to its nonpolynomial character this product induces nonlocal interaction effects in NC field theories.
The star-product is associative and has the important property of cyclicity upon spacetime integration:
\begin{eqnarray}
\int d^4 x \; f(x)*g(x)=\int d^4 x \; g(x)*f(x)=\int d^4 x \; f(x)g(x).
\label{cyclicity}
\end{eqnarray}

In the case of canonical NC spacetime the Lagrangian is constructed from products and derivatives of the NC operator fields. Then the action is defined as the trace of the Lagrangian. In terms of the Weyl symbols this corresponds to the spacetime integral of the Lagrangian, which can be written using the star-product. In terms of the above NC spacetime algebra, this trace corresponds to trace over a subspace where the operator $\hat \theta$ has a fixed eigenvalue. To obtain a Lorentz invariant formulation, the authors of \cite{Carlson:2002wj} generalized the trace by replacing the spacetime integral with a Lorentz invariant integration over all values of both coordinates $x$ and $\theta$:
\begin{eqnarray}
Tr \hat \phi (\hat x , \hat \theta) = \int d^4 x\;d^6\theta\; W(\theta)  \phi(x,\theta),
\label{}
\end{eqnarray}
where $W(\theta)$ is some Lorentz invariant weight function and $d^6\theta=d\theta^{01}d\theta^{02}d\theta^{03}d\theta^{12}d\theta^{13}d\theta^{23}$ is a Lorentz invariant measure.
Note that the theory can now be thought of as a 4+6 dimensional theory as the parameters $\theta^{\mu\nu}$ behave as six additional coordinates. However, it is to be assumed that there are no derivatives with respect to these extra coordinates and thus no propagation in the $\theta$-direction. Thus there is no need for a compactification of the extra dimensions. This also allows us to consider the fields as functions of $x$ only \footnote{This assumption may not be legitimate in the case of NC gauge theory, where the gauge transformations necessarily induce a $\theta$-dependence in the fields. In this paper we restrict the discussion to non gauge theories}.

As the NC parameter is now integrated over, noncommutativity enters through the choice of the weight function $W$. The commutative limit is equivalent to choosing $W$ as the delta function concentrated at the origin of the $\theta$-space, while the canonical case corresponds to a delta function in a nonzero point.
In the case of Lorentz invariant formulation one can impose some reasonable restrictions on the form of $W$. The weight function should be an even function of $\theta^{\mu\nu}$. We also impose the normalization
\begin{eqnarray}
\int d^6\theta\;W(\theta)=1,
\label{}
\end{eqnarray}
and $W$ is assumed to vanish rapidly enough at infinity in order to make all the needed integrals finite.

In perturbative calculations in Lorentz invariant NC field theory, one typically has to deal with phase factors of the form:
\begin{eqnarray}
V(k,p)=\int d\theta W(\theta) e^{-ik\wedge p},\;\;\;\;k\wedge p =\theta^{\mu\nu}k_\mu p_\nu,
\label{}
\end{eqnarray}
which appear in each vertex.
As was argued in \cite{Morita:2002cv,Morita:2003vt} by choosing the weight function in a Gaussian form, the Lorentz invariant integration brings such phase factors into the form of an exponential damping factor:
\begin{eqnarray}
V(k,p)=e^{-a^4(k^2p^2-(k\cdot p)^2)/4}.
\label{exponent}
\end{eqnarray} 
Here the parameter $a$ is related to the scale of noncommutativity, that is determined by the exact form of the weight function
\begin{eqnarray}
a^{4}=\langle\theta^2\rangle=\int d^6\theta \;W(\theta) \;\theta^{\mu\nu}\theta_{\mu\nu}.
\label{}
\end{eqnarray} 
The expression (\ref{exponent}) will be useful in our later calculations.

Finally we remark that the case of rotational invariant noncommutativity considered in \cite{Doplicher:1994zv,Doplicher:1994tu} corresponds to choosing the weight function to be concentrated on the rotational orbit of some chosen reference value $\theta_0^{\mu\nu}$:
\begin{eqnarray}
\int d^6 \theta W(\theta)\rightarrow \int_{\Sigma^{(1)}}d^6 \theta.
\label{}
\end{eqnarray}
Any nonzero choice for $\theta_0^{\mu\nu}$ obviously breaks Lorentz invariance, but the integration restores rotational invariance.


\sect{Causality in Lorentz invariant NC QFT}

To study causality we calculate a matrix element of an equal time commutator of an observable in NC theory. 
In NC spacetime a local observable that is in general a product of fields should be constructed using the noncommutative product. Thus even in the case of free fields noncommutativity enters through the definition of local observables even though the action is equivalent to the action in commutative spacetime due to the property (\ref{cyclicity}). In order to prove acausality it is then enough to demonstrate it for noncommutative observables in the free case.

In \cite{Chaichian:2002vw} the authors considered causality in the canonical NC spacetime by calculating the matrix element:
\begin{eqnarray}
\mathcal{M} =\bra{0}\left[:\phi(x)*\phi(x):,:\phi(y)*\phi(y):\right]_{x^0=y^0}\ket{p,p'}.
\label{canonical}
\end{eqnarray}
Here normal ordering is imposed for simplicity.
In a Lorentz invariant theory causality requires this equal-time commutator to vanish for all nonzero values of $\mathbf{x}-\mathbf{y}$.
In the case of canonical NC theory the vanishing of (\ref{canonical}) should occur only outside of the light wedge, i.e. for the nonzero values of 
$(x_2-y_2)^2+(x_3-y_3)^2$ in the case where only $\theta^{23}$ is nonzero.
The free field can be expanded in terms of creation and annihilation operators as in usual commutative theory:
\begin{eqnarray}
\phi(x)&=&\int\frac{d^3 k}{(2\pi)^3}\frac{1}{\sqrt{2\omega_\mathbf{k}}}\left(a_\mathbf{k}e^{-ikx}+a^\dagger_\mathbf{k}e^{ikx}\right)\bigg{|}_{k_0=\omega_\mathbf{k}},\nonumber\\
\left[ a_\mathbf{k} , a^\dagger_\mathbf{k'} \right]&=&(2\pi)^3\delta^3(\mathbf{k}-\mathbf{k}').\nonumber
\label{fields}
\end{eqnarray}
Inserting the expansion into (\ref{canonical}) we obtain 
\begin{eqnarray}
\mathcal{M}&=& -\frac{2i}{(2\pi)^6}\frac{1}{\sqrt{(\omega_\mathbf{p}\omega_{\mathbf{p}'})}}(e^{-ip'x-ipy}+e^{-ipx-ip'y})\nonumber\\
&\times&\int\frac{d^3k}{\omega_\mathbf{k}} \sin[\mathbf{k}(\mathbf{x}-\mathbf{y})]\;\cos\left(\frac{1}{2} k\wedge p \right)\;\cos\left(\frac{1}{2} k\wedge p'\right).
\label{}
\end{eqnarray}
Here $\omega_\mathbf{k}=\sqrt{\mathbf{k}^2+m^2}$. 
Obviously the r.h.s. is nonzero only when $\theta^{0i}\not=0$.

In the Lorentz invariant case the corresponding quantity is
\begin{eqnarray}
\mathcal{M}_\mathrm{LI}=\bra{0}\left[\int d\theta_1:\phi(x)*_1\phi(x):,\int d\theta_2:\phi(y)*_2\phi(y):\right]_{x^0=y^0}\ket{p,p'},
\label{}
\end{eqnarray}
where $*_i$ is the star-product corresponding the integration variable $\theta_i^{\mu\nu}$ and $d\theta_i:=d^6 \theta_i\;W(\theta_i)$. 
It is then easy to see that this produces the result of the canonical case with appropriate $\theta$-integrations added:

\begin{eqnarray}
\mathcal{M}_\mathrm{LI}&=&-\frac{2i}{(2\pi)^6}\frac{1}{\sqrt{(\omega_\mathbf{p}\omega_\mathbf{p}')}}(e^{-ip'x-ipy}+e^{-ipx-ip'y})\nonumber\\
&\times&\int\frac{d^3k}{\omega_\mathbf{k}} \left\{\sin[\mathbf{k}(\mathbf{x}-\mathbf{y})]\; \int d\theta_1\cos\left(\frac{1}{2} k\wedge_1 p \right)\;\int d\theta_2\cos\left(\frac{1}{2}k \wedge_2 p'\right)\right\}.
\label{causalityint}
\end{eqnarray}
We can see the acausality clearly by assuming a Gaussian form for the weight function. Then we can use (\ref{exponent}) to write the $\theta$-integrals as:
\begin{eqnarray}
&\displaystyle\int & d \theta \cos\left(\frac{k\wedge q}{2}\right)\nonumber\\
=&\exp&\left[-\frac{a^4}{16}(-q_0^2\mathbf{k}^2-k_0^2\mathbf{q}^2+\mathbf{k}^2\mathbf{q}^2+2k_0q_0\mathbf{k}\cdot \mathbf{q}-(\mathbf{k}\cdot\mathbf{q})^2)\right].
\label{cos}
\end{eqnarray}
The fourth term in the exponent prevents this from being an even function of $\mathbf k$.
Applying (\ref{cos}) in (\ref{causalityint}) one sees that that the function multiplying the $\sin[\mathbf{k}(\mathbf{x}-\mathbf{y})]$ in the integrand is even in $\mathbf{k}$ only if $k_0p_0\mathbf{k}\cdot\mathbf{p}+k_0p'_0\mathbf{k}\cdot\mathbf{p'}$ vanishes.
Thus the matrix element vanishes for nonzero $\mathbf{x}-\mathbf{y}$ only if $p_0\mathbf{p}+p'_0\mathbf{p}'=0$. This is satisfied if the two-particle state has zero total momentum and both particles are on-shell, but not in general and this is enough to show that the commutator is a nonzero operator.
Due to the Lorentz invariant integration nonlocality is infinite in all directions and one cannot recover the light-wedge causality condition that one has in the canonical case.

Finally, we make a comment on the rotational invariant noncommutativity advocated in \cite{Doplicher:1994zv,Doplicher:1994tu}. In this case 
the Lorentz invariant
$\theta$-integrations in (\ref{causalityint}) should be replaced with integrations over the rotational orbits $\Sigma^{(1)}$ of some reference value $\theta_0$. Then, in a Lorentz frame where the time-space part of $\theta_0$ is nonvanishing causality is lost.


\sect{Covariant versus time-ordered perturbation theory}

Next we will address the issue of perturbation theory in the Lorentz invariant approach. 
As it has been shown in \cite{Morita:2003vt} unitarity seems to hold in quantum field theories in Lorentz invariant NC spacetime
despite the fact that due to the Lorentz invariant integration also time-space noncommutativity is present. 
This result was obtained in the lowest order of
perturbation using covariant Feynman rules. 
On the other hand in the time-ordered perturbation theory unitarity is manifest and thus it is interesting to see whether in the Lorentz invariant NC theory the time-ordered and covariant formulations actually coincide as in ordinary commutative field theory. 

To compare the two approaches we simply write down the amplitudes for two-by-two particle scattering in $\phi_*^3$ theory, with the action,
\begin{eqnarray}
S=\int d^6\theta\; d^4 x\ \left(\frac{1}{2}(\partial_\mu \phi)^2-\frac{1}{2}m^2\phi^2-\frac{\lambda}{3!}\phi*\phi*\phi \right). 
\label{}
\end{eqnarray} 
The rules for calculating scattering diagrams in NC TOPT were derived and listed in \cite{Liao:2002xc}. The only difference in the Lorentz invariant case is that there is a $\theta$-integration included in each vertex. Using these rules we obtain for the tree level amplitude (up to a constant factor):

\begin{eqnarray}
\mathcal M&=& \int d\theta_1d\theta_2\;\delta^4(k_1+k_2-k_3-k_4)\frac{1}{\omega_{\mathbf k_1+\mathbf k_2}} \nonumber\\
&\times & \left\{(k_{1,0}+k_{2,0}-\omega_{\mathbf k_1+\mathbf k_2}+i\epsilon )^{-1}\Sigma_{sym.}\cos_{\theta_1}\left(k_1^{(+)},k_2^{(+)},-p_{\mathbf k_1+\mathbf k_2}^{(-)}\right)\right.\nonumber\\
&& \times \Sigma_{sym.}\cos_{\theta_2}\left(-k_3^{(-)},-k_4^{(-)},p_{\mathbf k_1+\mathbf k_2}^{(+)}\right)\\
&& + (-k_{1,0}-k_{2,0}-\omega_{\mathbf k_1+\mathbf k_2}+i\epsilon )^{-1} \Sigma_{sym.}\cos_{\theta_1}\left(k_1^{(+)},k_2^{(+)},-p_{\mathbf k_1+\mathbf k_2}^{(+)}\right)\nonumber \\
&&\left.\times\Sigma_{sym.}\cos_{\theta_2}\left(-k_3^{(-)},-k_4^{(-)},p_{\mathbf k_1+\mathbf k_2}^{(-)}\right)\right\}.\nonumber
\label{TOPT}
\end{eqnarray}
Here $k_i$ are the external momenta and we have used the notation
\begin{eqnarray}
k^{(\pm )}_i&=&(\pm k_{i,0},\mathbf k_i), \nonumber \\
p^{(\pm )}_{\mathbf k_1+\mathbf k_2}&=&(\pm \omega_{\mathbf k_1+\mathbf k_2}, \mathbf k_1+\mathbf k_2),\\ 
(a,b,c)&=&a\wedge b+a\wedge c+b\wedge c. \nonumber
\label{notation}
\end{eqnarray}
The wedge products in the argument of $\cos_{\theta_{i}}$ are defined with respect to $\theta_{i},\;i=1,2$  respectively and $\Sigma_{sym.}$ implies symmetrization over the particles in the vertex.
The corresponding expression for the amplitude with the covariant Feynman rules is up to a constant factor,
\begin{eqnarray}
\mathcal{M}&=&\int d\theta_1 d\theta_2 \left\{\frac{1}{(k_1+k_2)^2-m^2+i\epsilon}\right.\nonumber\\
&&\cos (k_1\wedge_1 k_2)\cos (k_3\wedge_2 k_4) \delta^4(k_1+k_2-k_3-k_4)\bigg\}.
\label{covariant}
\end{eqnarray}

The usual commutative analogues of (4.2) 
and (4.4) do not have the cosine factors and integrations over $\theta_i$. In that case the two terms in the TOPT expression combine trivially to the form of the covariant expression. In order to compare the NC expressions we first note that 
\begin{eqnarray}
\cos\left(a,b,c\right)=\cos\left(a\wedge b+a\wedge c+b\wedge c\right)=\cos\left((a+b)\wedge(a+2ab+c)\right).
\label{}
\end{eqnarray}
Then we can once again use the Gaussian expression (\ref{exponent}) to obtain an exponential form for the NC vertex factor
\begin{eqnarray}
&\int & d\theta \cos(k_1^{(+)},k_2^{(+)},-p^{(-)}_{\mathbf k_1+\mathbf k_2})\nonumber\\
&=&\exp\left\{-\frac{a^4}{4}\bigg[(k_1+k_2)^2(k_1+2k_2-p^{(-)}_{\mathbf k_1+\mathbf k_2})^2\right.\\
&-&\left.\left.\left((k_1+k_2)^\mu(k_1+2k_2-p^{(-)}_{\mathbf k_1+\mathbf k_2})_\mu\right)^2\right]\right\},\nonumber
\label{exptopt}
\end{eqnarray}
and similar expressions for the other vertex factors in (4.2).
The exponents contain zeroth components of the internal momentum $p^{(\pm)}_{\mathbf k_1+\mathbf k_2}$, which are of the form $\omega_{\mathbf k_1+\mathbf k_2}$ and which do not cancel after combining all the vertex factors in (4.2). Corresponding terms do not arise in the $\theta$-integrals of the cosine factors in (\ref{covariant}) and thus (4.2)
can not reduce to the covariant form (\ref{covariant}). This is sufficient for demonstrating that the time-ordered and covariant formulations of perturbation theory are inequivalent in the Lorentz invariant NC theory. Note also that if the time components of all the momenta in the vertex factors after the $\theta$-integrations are ignored, i.e. if there were no contributions from time-space components of $\theta$, the TOPT expression for the amplitude would reduce to the covariant expression similarly to the case of canonical noncommutativity with $\theta^{0i}=0$.


\sect{Discussion and conclusion}
In this paper we have considered basic properties of Lorentz invariant NC field theory. The Lorentz invariance allows for the possibility that the ordinary light-cone causality condition holds. However, we showed that despite the manifest Lorentz invariance such theories are acausal. It is clear that the UV/IR mixing effect is intimately related to the acausality in NC field theory. The phase factors that lead to violation of causality provide UV cutoffs in loop diagrams. As the external momentum approaches zero the damping effect disappears and the UV divergence reappears as an IR singularity.  
This mixing between short and long distance degrees of freedom can be understood in terms of acausality. Due to infinite propagation speed all distance scales are correlated. As we have seen, in the Lorentz invariant case causality is violated in all directions and thus the UV/IR mixing is expected to appear.
Due to the $\theta$-integration the oscillating phase factors in loop diagrams get replaced by Lorentz invariant Gaussian damping factors. Indeed these again lead to IR singularity as the external momentum goes to zero. However, it was argued in \cite{Morita:2002cv} that the problem may be avoided by a suitably chosen IR limit under which $a$ goes to zero with the external momentum -- an argument which does not work without Lorentz invariance. 
In any case lack of causality is a problem that cannot be dismissed and it is necessary to find a way to restore the light-cone causality in order to achieve a consistent NC field theory. 

We have also found that the issue of perturbation theory in Lorentz invariant NC field theory is ambiguous since the time-ordered and covariant formulations do not coincide. In \cite{Morita:2003vt} unitarity was shown to hold in $\phi_*^3$-theory in the lowest order in covariant perturbation theory.
In the light of our result it is reasonable to suspect that the result of \cite{Morita:2003vt} does not
hold in a more general theory or at higher orders, and to retain unitarity one should use TOPT.
On the other hand, in the case of canonical NC theory the TOPT formulation does not cure the problems arising from time-space noncommutativity completely since in the TOPT formulation the Ward identities are violated in gauge theories \cite{Ohl:2003zd}. It would be interesting to see whether this still holds after restoring the Lorentz symmetry and to investigate in more detail the differences between the time-ordered and covariant approaches to perturbation theory.

Finally, we mention that in the case of NC gauge theories one could avoid these problems at least superficially by using the Seiberg-Witten map. Taking any finite order in the expansion in $\theta$, one avoids problems arising from the nonlocal character of the star-product. However, the expansion in the NC parameter may miss important aspects of NC physics and thus it would be
preferable to obtain a consistent formulation of Lorentz invariant NC theory in the full star-product formalism.

%
%


\vspace*{10mm}
\begin{center}
{\bf Acknowledgements}
\end{center}
The author is grateful to Masato Arai, Masud Chaichian, Anca Tureanu and Nobuhiro Uekusa for useful discussions during the preparation of this paper. 

%
%

\small{

\end{document}